Report on the ESO workshop

# LISA 10 — Library and Information Services in Astronomy

held at ESO Vitacura, Santiago, Chile, 3–7 November 2025

Leslie Kiefer[1]
Uta Grothkopf[1]
Silvia Meakins[1]
Mahmut Eren[1]

[1] ESO

**The 10th LISA conference (LISA 10) took place at ESO Chile in early November 2025. The theme, Research Equity and Access in the Age of AI, highlighted the rapidly evolving landscape of scholarly communication in astronomy. Artificial intelligence, open access mandates, and expanding data archives are transforming scientific practice and information in an increasingly interdisciplinary research environment. Through presentations, posters, and discussions, LISA 10 explored how these developments are redefining the management of astronomy information work and the role of librarians.**

## Introduction

LISA 10[1] was the latest in the LISA (Library and Information Services in Astronomy)[2] series of international community-driven conferences that bring together astronomy librarians, information professionals, data specialists, archivists, documentalists, publishers and scientists to discuss issues relevant to information management in astronomy and related sciences. The series began in 1988 and has been held every three to four years since LISA II (held at ESO in Garching in 1995).

Astronomy librarians traditionally have been a small but global and yet well-connected community. LISA conferences have always been special as they are the foremost international meeting in our profession, enjoying high visibility and recognition (Figure 1).

## Day 1: AI/ML, the UAT and the evolving role of libraries

Astronomy, like many other sciences, is seeing artificial intelligence (AI) and machine learning (ML) advance rapidly. This was reflected by the first conference session which showed how ML is entering scientific workflows, from identifying features in astronomical images to detecting the use of observational data in scientific papers. These projects illustrated AI's potential to support data curators. Librarians and publishers addressed the ethical aspects of AI, stressing the need for policies to guide responsible use of AI- and ML-based tools.

Several presentations examined the Unified Astronomy Thesaurus (UAT), its use as a training tool and its integration into systems such as the ADS/SciX database. Speakers emphasised that shared, consistent metadata standards remain essential even as automated indexing expands.

In an invited talk, Rafael Castillo and Cristián Calabrano of the University of Chile described the university's emerging open science ecosystem, shaped by years of policy development and practical experience to create a science environment that is valued by researchers.

The day concluded with a look at how astronomy libraries may evolve in a rapidly changing society, a theme that recurred throughout the conference.

Figure 1. LISA 10 conference photo at ESO Vitacura.

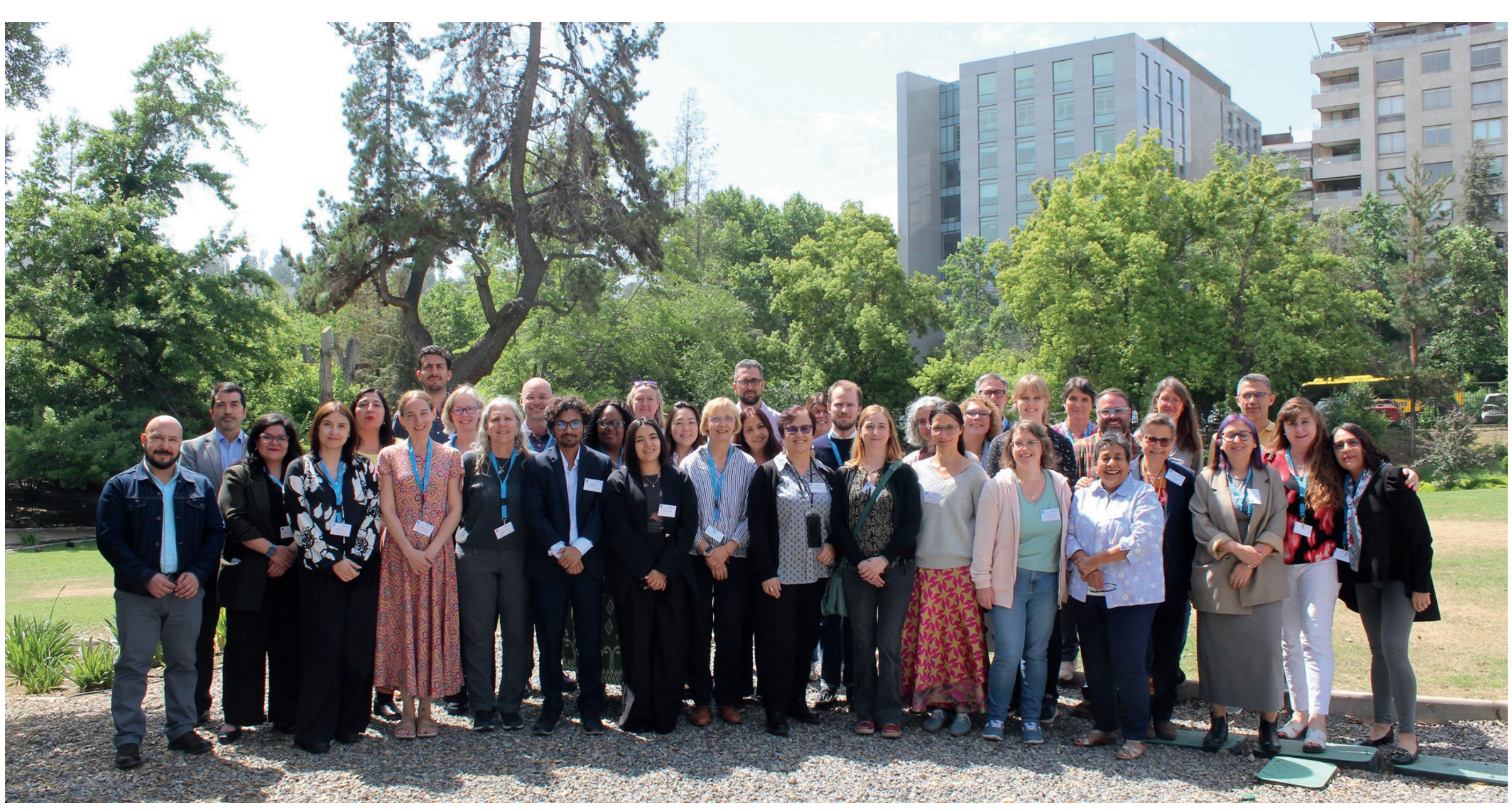

## Day 2: Open access and research equity

Day 2 discussions focused on open access (OA), open science and the future of publishing. Astronomy has long been a leader in open data practices and manuscript sharing, but journal publishing in other sciences has taken different directions, creating a complex publishing landscape shaped by publishers’ business models, funder mandates and community-led dissemination.

Speakers presented a range of OA models, including the Astronomy & Astrophysics (A&A) adaptation of the collaborative Subscribe to Open (S2O) model, and the Open Journal of Astrophysics (OJAp) which relies on the arXiv platform to operate an overlay journal without author-facing Article Processing Charges (APCs). Presenters also discussed OA trends in India, the use of transformative agreements to fulfill national OA mandates in Italy and how stakeholders can shape the future of scientific publishing. While the principles of openness enjoy broad support, achieving an equitable, transparent and sustainable publishing ecosystem remains challenging.

A highlight was a lively presentation by invited speaker María Soledad Bravo-Marchant from the Consorcio para el Acceso a la Información Científica Electrónica (CINCEL), who offered a comprehensive look at SciELO Chile, an open access journal platform managed by the Agencia Nacional de Investigación y Desarrollo (ANID). Her talk illustrated how national initiatives can enable sustainable platforms for quality OA publishing, especially in contexts where commercial APC-based models are prohibitive.

A set of presentations addressed the importance of data linking and best practices for data publication and parallels between earth sciences and astronomy, underscoring the importance of interoperability and standardised citation. Additionial talks highlighted very successful virtual exhibitions, initiatives to improve access to astronomical knowledge and national strategies for open science training in Chile.

In the late afternoon participants visited the historic Manuel Foster Observatory on San Cristóbal Hill (Figure 2). Originally built by Lick Observatory astronomers and now a national monument managed by the Pontificia Universidad Católica de Chile, the site provided an excellent link between traditional and contemporary astronomy — a perfect end to the day.

## Day 3: Open science, systems, and data

The third day of LISA 10 examined the infrastructures necessary to make open science a reality. Starting from bibliometric analyses and impact studies, the focus turned to data archives, with presentations from NASA’s Mikulski Archive for Space Telescopes (MAST), the Strasbourg astronomical Data Center (CDS) and the NASA/IPAC Extragalactic Database (NED) at Caltech. These talks showcased the complex and meticulous work carried out by data specialists and documentalists, applying their expertise in data curation, linking, standardisation and long-term preservation.

Several talks highlighted the growing demands of data stewardship. As astronomy generates increasingly complex datasets, the roles of librarians and documentalists have evolved from custodians to connectors responsible for identifying, contextualising and shaping the standards used across the core services of our profession. A joint presentation of colleagues from the CDS and NED demonstrated the extensive efforts required to harvest data from astronomical papers; the talk also illustrated the logistical challenges for the conference organisers in scheduling a session that accommodated speakers in Paris and California, with an online audience extending towards India and beyond.

Invited speaker Chris Erdmann from SciLife Lab in Sweden offered a cross-disciplinary perspective. Based on his extensive experience in both astronomy and the life sciences, he compared Findability, Accessibility, Interoperability and Reusability (FAIR) tools, platforms and implementation strategies in the two fields, emphasising that astronomy’s long-standing open data culture provides a strong basis for open science principles that extend beyond data to include software, research methods, infrastructures and the application of responsible metrics.

The day concluded with the conference dinner at a spectacular location in the Cajón del Maipo, Roan Jasé, providing home-cooked food, a tour of the site’s educational resources, including a 12-inch refractor telescope used for school classes and other visitors, and ample opportunities for conversations among participants and hosts.

## Day 4: Stewarding astronomical knowledge and the changing roles of libraries

The final day began with invited speaker Álvaro López, who provided an overview of the national open science efforts in Chile, connecting high-level policy with the practical challenges faced by librarians, educators and researchers. His talk brought together many of the conference’s core themes, including openness, infrastructure, and the importance of equity.

Case studies from the Inter-University Centre for Astronomy and Astrophysics (IUCAA) in India as well as UC Berkeley and the U.S. Naval Observatory in the USA illustrated how astronomy libraries worldwide are navigating evolving landscapes of print and digital collections, constrained staffing resources and changing user expectations. Additional talks focused on digitising centuries-old scientific correspondence, recovering the names and contributions of women in scientific archives, revising collection development in an increasingly electronic environment, and strategies to establish legacy collections. These presentations highlighted both the adaptability of information professionals, their creativity in reaching diverse user groups and their attention to historical collections at a time when the establishment of new facilities and the application of cutting-edge technology often take precedence.

Speakers also addressed the ever-changing role of astronomy librarians. Several presentations throughout the conference made it clear that the profession is not experiencing a revolution, but rather an evolution, in which we apply the tools of our trade, along with our skills and expertise, in the digital era.

The conference concluded with a historical retrospective on the LISA series by colleagues who have participated in (almost) all meetings since the first one in 1988 (not called LISA I as no one then knew there would ever be a second). Over time, LISA conferences have grown to become a globally recognised forum which continues to strengthen the community's mission: to collaborate and unite human expertise to advance information management in astronomy.

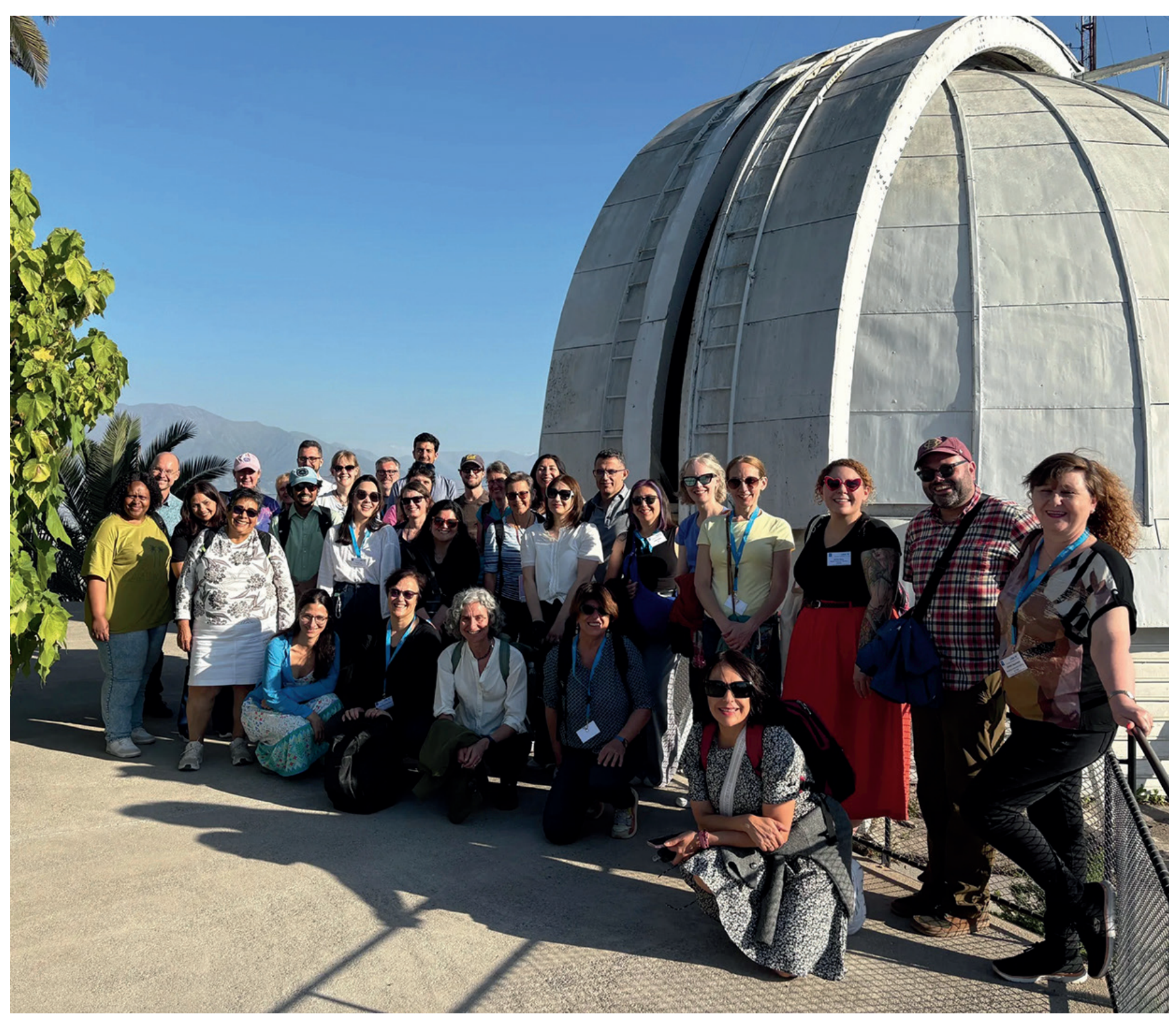

Figure 2. Excursion to Manuel Foster Observatory, San Cristóbal Hill, Santiago de Chile

Following the core conference, some participants enjoyed a self-paid trip to visit ESO's Paranal Observatory, which allowed them a glimpse of the Extremely Large Telescope from a distance. The combination of cutting-edge technology and the seemingly endless Atacama Desert left everyone deeply impressed!

## Looking ahead

LISA 10 provided a lively picture of a profession in transition. AI promises new capabilities, but requires ethical frameworks. Data systems are growing in complexity, yet the expertise needed to curate and connect information is often hidden behind user-friendly interfaces. Open access publishing offers the possibility of universal research participation, but inadequately formulated policies and mandates have exacerbated existing inequalities. Today's science ecosystem requires close collaboration among librarians, data specialists, publishers and researchers. LISA 10 confirmed that even in an age of automation and global digital infrastructures, the core of scholarly communication continues to be defined by humans.

At the end of LISA 10, a group of colleagues volunteered to host the next event. Such an early commitment had not happened before in the history of the conference series and it showed the importance and value the meetings continue to have for the community. Long live LISA!

## Demographics

Traditionally, LISA conferences have enjoyed participation from colleagues around the globe. LISA 10 welcomed 80 online (52.5%) and in-person (47.5%) participants from 16 countries, including Latin America (Chile, Mexico), Europe (Finland, France, Germany, Ireland, Italy, Norway, Spain, Sweden, The Netherlands, UK), North America (Canada, US) as well as South Africa and India. The Scientific Organising Committee invited four speakers (one female, three male) from Chile and Sweden. As can be expected in our profession, a large fraction of the presentations was delivered by female speakers (almost 70%), while male speakers presented approximately 30% of the talks.

## Acknowledgements

This conference would not have been possible without the scientific organising committee chairperson, Jenny Novacescu, formerly at the Space Telescope Science Institute, who could not attend LISA 10 for personal reasons. A heartfelt Thank You to Jenny for everything she has done to make LISA 10 happen!

In order to ensure full inclusion, in particular of colleagues from our host country Chile, LISA 10 provided a simultaneous interpretation service that ensured that presentations in English or Spanish were understandable to non-native speakers of either language. We would like to thank the interpreters for their excellent work.

The organisers gratefully acknowledge the generous support of the conference sponsors: American Association for the Advancement of Science, EDP Sciences, ESO, IOP Publishing, Open Fifth, the Royal Astronomical Society and the SPIE Digital Library.

The LISA 10 conference proceedings (presentation slides and posters) are available via a dedicated collection at Zenodo[3].

## Links

[1] LISA 10 homepage: https://www.eso.org/sci/meetings/2025/LISA10.html
[2] LISA conferences: https://www.eso.org/sci/libraries/lisa.html
[3] LISA 10 presentations and posters: https://zenodo.org/communities/lisa10/